\documentclass[aps,prl,reprint,superscriptaddress,floatfix]{revtex4-2}
\usepackage{graphicx}
\usepackage{amsmath,amssymb,bm}
\usepackage{placeins}
\usepackage{xcolor}
\definecolor{BLUE}{rgb}{0,0,1}  
\usepackage[colorlinks=true,linkcolor=blue,citecolor=blue,urlcolor=blue]{hyperref}

\begin{document}

\title{Which pulse maximizes resonant nonlinear conversion?}

\author{Alex Krasnok}
\affiliation{Department of Electrical and Computer Engineering,
Florida International University, Miami, Florida 33174, USA}

\date{August 19, 2026}

\begin{abstract}
At fixed pulse energy, which drive waveform extracts the most $n$th-order nonlinear
conversion from a resonator ($n=2$ for second harmonic)? A short pulse couples poorly
to a narrow resonance, a long one dilutes its energy, and no linear rule fixes the
compromise. We solve the problem exactly for a single mode of amplitude decay rate
$\kappa$. Eliminating the drive turns fixed incident energy into a constraint on the
stored field alone, and the optimization becomes a sharp Gagliardo--Nirenberg
inequality whose extremal is the ground-state soliton of the nonlinear Schr\"odinger
equation. The optimal stored field is $\mathrm{sech}^{1/(n-1)}[(n-1)\kappa t]$,
sustained by an asymmetric input that rises as $e^{\kappa t}$ and falls as
$e^{-(2n-1)\kappa t}$; the largest converted energy follows in closed form. A rising
exponential, the time-reversal recipe, retains at most $79.0\%$ of the bound at $n=2$
and $2/e$ at large $n$; a two-rate pulse retains above $97\%$. Critical coupling
generalizes to $n$-fold overcoupling, with optimal input coupling $n$ times the
intrinsic loss rate. The bound applies from microrings to superconducting circuits and
caps the per-pulse brightness of broadband photon-pair sources.
\end{abstract}

\maketitle

Resonators make nonlinear optics work at modest power by storing the pump. In the
linear case two rules settle how best to load one. Critical coupling matches the input
coupling rate to the intrinsic loss rate and cancels reflection under steady driving.
Time-reversed loading drives with an exponential rising at the field amplitude decay
rate $\kappa$, which maximizes the stored energy at one
instant~\cite{Stobinska2009,Heugel2010,Bader2013}. A nonlinear device asks a different
question. In $n$th-order conversion---$n=2$ for second-harmonic generation, $n=3$ for
third---the generated power follows the $n$th power of the stored intensity, so the
yield is the time integral $\int|a|^{2n}dt$, where $|a|^2$ is the stored energy.
Shortening the pulse raises the peak but spills energy outside the linewidth;
stretching it dilutes the intensity the nonlinearity rewards. Neither linear rule fixes
the compromise.

Optimization has so far run inside chosen families: pulse duration in resonant harmonic
generation~\cite{Nikitin2025,Shcherbakov2019,Bijloo2025,Franceschini2024}, spectral
phase at fixed energy~\cite{Meshulach1998,Meron2025}, rising exponentials and their
complex-frequency generalizations for the linear loading
problem~\cite{Baranov2017,Radi2020,Hinney2024,Kim2025Review,Xue2026,KrasnokSeletskiy2026},
and modulation profiles of time-varying cavities whose observable is
linear~\cite{CortesHerrera2022}. Separately, a large body of work bounds what a
photonic \emph{structure} can do, limiting scattering, absorption, and response by
geometry and material~\cite{Molesky2018,Chao2022}. The matching statement for the
\emph{drive}---the largest nonlinear yield any pulse of given energy can produce---has
been missing.

Here we supply it in closed form for a single resonance. Fixing the incident energy
fixes a quadratic norm of the stored field, and maximizing $\int|a|^{2n}dt$ under that
constraint is a sharp functional inequality whose extremal is a soliton: the drive that
best exploits a lossy \emph{linear} resonator builds the ground state of a lossless
\emph{nonlinear} wave equation. The bound holds over all square-integrable inputs and
is attained, so it is an absolute yardstick rather than the best member of a family,
and the unknown nonlinear coefficient cancels between two waveforms of equal energy,
making the ratios below testable. Two design rules follow: the input coupling should be
$n$ times the intrinsic loss rate, and the best rising-exponential rate is
$(n-1)\kappa/n$, which still forfeits $21\%$--$26\%$ of the yield; a continuous
two-rate pulse recovers more than $97\%$ for integer $2\le n\le20$.

\textit{Model and bound.---}One resonant mode driven through one port has complex
amplitude $a(t)$, normalized so that $|a|^2$ is the stored energy, obeying the
coupled-mode equation~\cite{Haus1984,Fan2003}
\begin{equation}
 \dot a=-(\kappa+i\Delta)a+\sqrt{2\kappa_e}\,s(t),
 \qquad \kappa=\kappa_e+\kappa_i ,
 \label{eq:cmt}
\end{equation}
where $s(t)$ is the incident field with $|s|^2$ the incident power, $\kappa_e$ the
decay rate into the driven port, $\kappa_i$ the sum of intrinsic loss and all undriven
ports, and $\Delta$ the detuning of the drive carrier from the resonance frequency
$\omega_0$. The loaded rate $\kappa$ is an \emph{amplitude} decay rate: stored energy
decays as $e^{-2\kappa t}$ and the loaded quality factor is $Q=\omega_0/2\kappa$.
Equation~(\ref{eq:cmt}) is also the input--output equation of a damped quantum
mode~\cite{Gardiner1985}; it holds whenever the resonance is spectrally isolated, the
port coupling is memoryless across the pulse bandwidth, and the mode profile does not
depend on stored energy~\cite{SM}. The incident energy $E=\int|s|^2dt$ is fixed
throughout.

\begin{figure*}[t]
\centering
\includegraphics[width=0.99\textwidth]{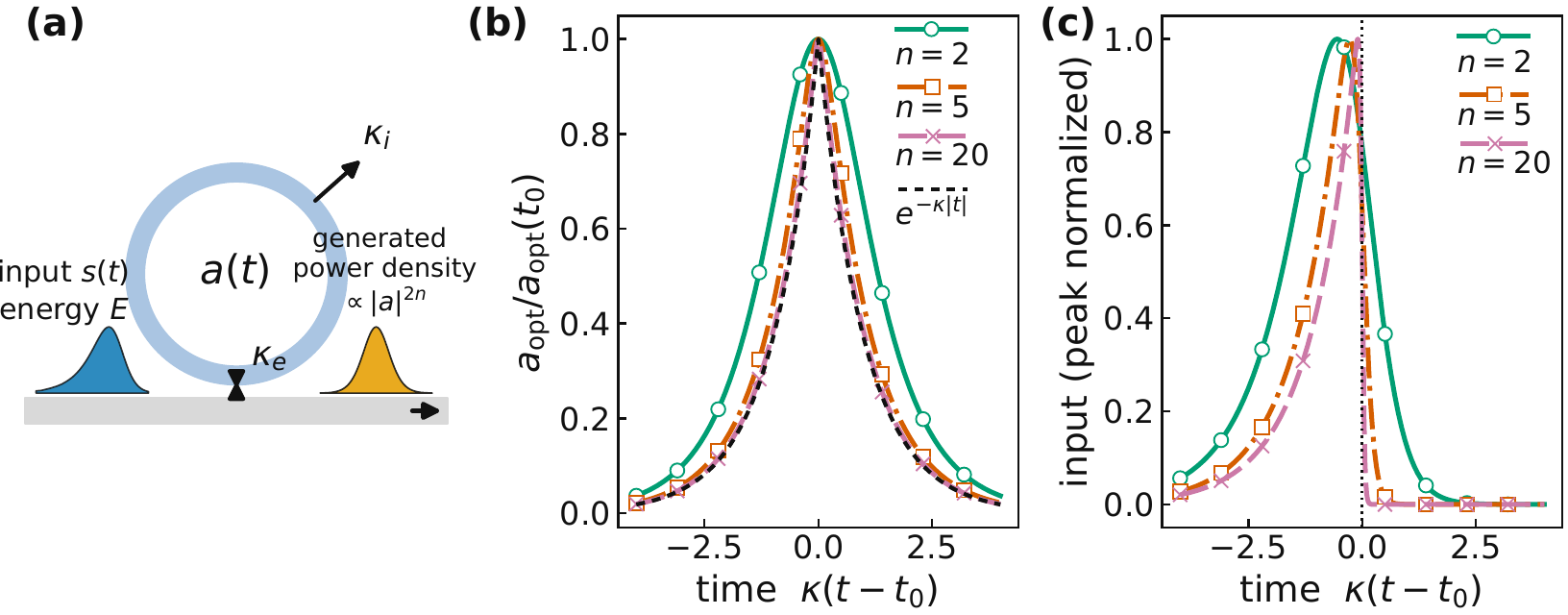}
\caption{Optimal driving of a single-mode resonator.
(a) A pulse of fixed energy $E$ enters through one port (coupling rate $\kappa_e$,
intrinsic loss $\kappa_i$); the stored field $a(t)$ produces the broad-channel
generated-power density $\propto|a|^{2n}$.
(b) Peak-normalized optimal stored fields, Eq.~(\ref{eq:aopt}), for $n=2,5,20$
(solid circles, dash-dotted squares, and long-dashed crosses), narrowing toward the
black short-dashed limit $e^{-\kappa|t|}$.
(c) Corresponding peak-normalized inputs, Eq.~(\ref{eq:sopt}), with the same order
encoding.  All share the rising tail $e^{\kappa t}$, while the trailing taper
steepens with order.}
\label{fig:exact}
\end{figure*}

In the undepleted regime---only a small fraction of the pump converts---with a material
response fast compared with the pulse and a generated channel spectrally flat across
the emission, conditions relaxed below, the converted energy is a process-dependent
constant times the shape functional $U_n=\int|a|^{2n}dt$. Harmonic generation of order
$n$ has this form for integer $n$; the bound below holds for any real $n>1$.

The step that solves the problem is to eliminate the drive. Solving
Eq.~(\ref{eq:cmt}) for $s$ and integrating $|s|^2$, the cross term is a total
derivative of $|a|^2$ and integrates to zero, leaving on resonance
\begin{equation}
 E=\frac{1}{2\kappa_e}\int\left(|\dot a|^2+\kappa^2|a|^2\right)dt .
 \label{eq:energy}
\end{equation}
The constraint no longer mentions $s$: fixed incident energy is a fixed value of the
$H^1$ Sobolev norm of the stored field, the sum of its squared slope and its squared
amplitude weighted by $\kappa^2$.

Maximizing $\int|a|^{2n}dt$ at fixed $H^1$ norm is the sharp one-dimensional
Gagliardo--Nirenberg inequality~\cite{Weinstein1983,Kwong1989}, an interpolation
estimate that bounds a high power of a function by  {how much slope and
how much amplitude it is allowed}.
Its extremal---the profile that turns the inequality into an equality---is the
ground-state soliton of the stationary nonlinear Schr\"odinger equation, with $\kappa^2$
playing the mass term. Writing $\rho=\kappa_e/\kappa$ for the loading fraction and
defining
$D_n=\sqrt\pi\,\Gamma\!\left(1+\tfrac{1}{n-1}\right)/\Gamma\!\left(\tfrac12+\tfrac{1}{n-1}\right)$
and $\mathcal C_n=\tfrac{2}{n+1}\left(\tfrac{n+1}{n}\right)^nD_n^{1-n}$, the result is
\begin{equation}
 \boxed{\ U_n\leq\frac{\rho^nE^n}{\kappa}\,\mathcal C_n\ }
 \label{eq:bound}
\end{equation}
with equality, up to a time shift $t_0$ and a constant phase, only for
\begin{equation}
 a_{\rm opt}(t)=A_n\,\mathrm{sech}^{\frac{1}{n-1}}\!\left[(n-1)\kappa(t-t_0)\right],
 \label{eq:aopt}
\end{equation}
where $A_n^2=\rho E(n+1)/(nD_n)$, sustained by the input that
Eq.~(\ref{eq:cmt}) run backwards demands,
\begin{equation}
 s_{\rm opt}(t)=\frac{\kappa A_n}{\sqrt{2\kappa_e}}\,
 \mathrm{sech}^{\frac{1}{n-1}}(x)\left[1-\tanh x\right],
 \label{eq:sopt}
\end{equation}
with $x=(n-1)\kappa(t-t_0)$. For second-harmonic generation, $D_2=2$ and
$\mathcal C_2=3/4$, so $\int|a|^4dt\le3\rho^2E^2/(4\kappa)$. The sharp inequality
certifies global optimality and its equality cases give uniqueness~\cite{SM}. The
equality waveform has infinite temporal support, so finite-aperture pulses approach the
bound without attaining it. At $n=1$ no maximizer exists: ever longer pulses approach
the supremum $2\kappa_eE/\kappa^2$. A finite optimal duration is itself a nonlinear
effect.

\begin{figure*}[t]
\centering
\includegraphics[width=0.80\textwidth]{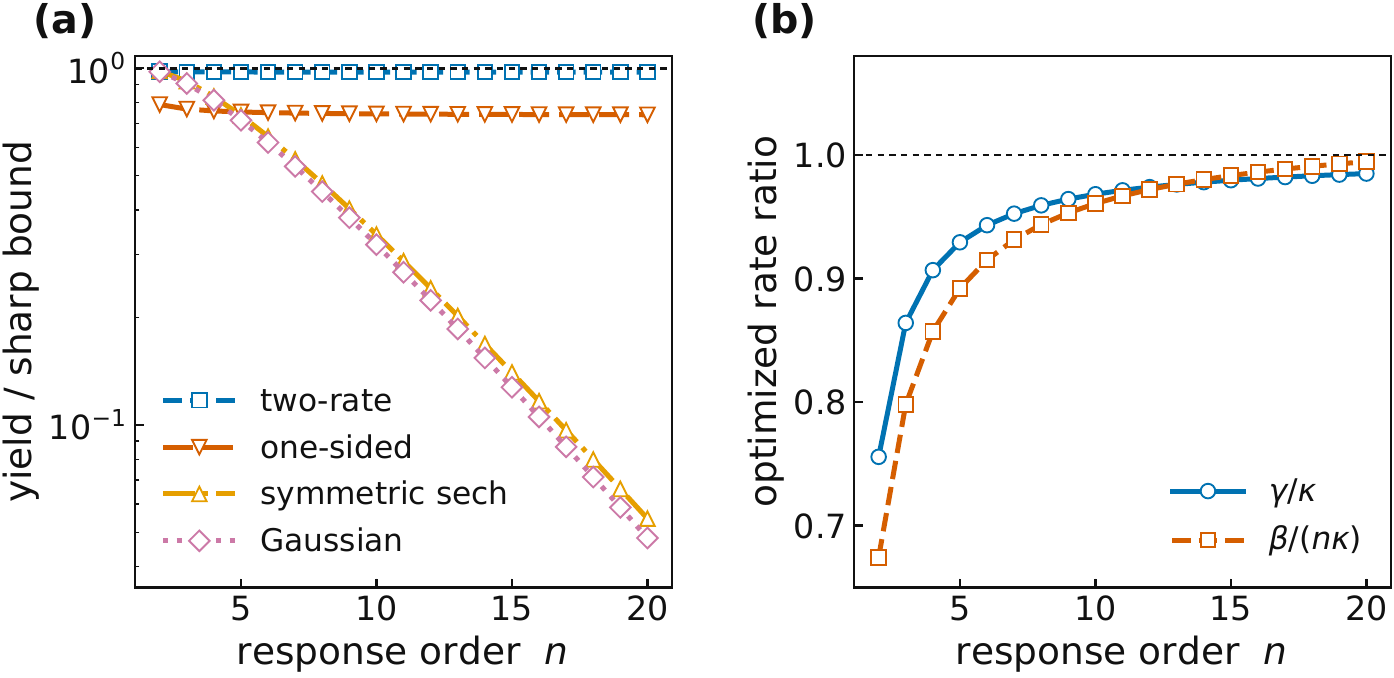}
\caption{Practical drives relative to the sharp bound.
(a) Retained yield for the optimized two-rate pulse (dashed squares), best one-sided
exponential (long-dashed downward triangles), width-optimized symmetric sech
(dash-dotted upward triangles), and Gaussian (dotted diamonds); the black short-dashed
line is the bound.
(b) Joint two-rate optima: solid circles show $\gamma/\kappa$ and dashed squares show
$\beta/(n\kappa)$, where $\gamma$ and $\beta$ are the leading-edge rise and
trailing-edge fall rates; the black short-dashed line marks unity.}
\label{fig:practical}
\end{figure*}

\textit{A soliton from a linear resonator.---}Equation~(\ref{eq:aopt}) is a soliton
profile. At $n=2$ it is exactly $A_2\,\mathrm{sech}[\kappa(t-t_0)]$, the fundamental
bright soliton of the cubic nonlinear Schr\"odinger equation---the pulse of fiber
optics~\cite{Zakharov1972,Hasegawa1973,Mollenauer1980} and of temporal cavity solitons
in driven Kerr resonators~\cite{Leo2010,Herr2014}. For other orders it
is the ground state of the same equation with a power nonlinearity of order $2n-1$,
sharpened from $\mathrm{sech}$ to
$\mathrm{sech}^{1/(n-1)}$~\cite{Weinstein1983,Kwong1989,KivsharAgrawal2003}.

 {The origin of the profile is different here.} A fiber or
microresonator soliton is
dynamical: a nonlinear medium supplies self-action, dispersion supplies spreading, and
their balance carries a pulse without reshaping it.  {Here the pump
resonator is linear, nothing propagates, and in the undepleted regime no nonlinearity
acts back on $a(t)$ at all.} The soliton appears because the
\emph{objective} is a high power of the field while the \emph{device} stays linear, so
the balance is struck by an optimization instead of by an equation of motion.

The optimal shape is therefore universal: it depends only on the response order and
the loaded linewidth. Neither the material nor the nonlinear
coefficient enters, and the pulse energy fixes the amplitude $A_n$ and nothing else, so
the same waveform is optimal for $\chi^{(2)}$ doubling in lithium niobate and for a
multiphoton process in a superconducting circuit. The soliton also survives the one
nonlinearity that does act on the pump, since a Kerr shift of the resonance leaves the
optimal envelope untouched and moves only its phase (below). Changing the measured
observable changes which soliton appears rather than whether one appears: a narrow
generated channel selects $\mathrm{sech}^{2/(n-2)}$, and multiwave mixing selects the
same family indexed by the total number of pump photons consumed. The soliton is the
\emph{stored} field; the input that builds it, Eq.~(\ref{eq:sopt}), is strongly
asymmetric, so a symmetric $\mathrm{sech}$ pulse from a mode-locked laser is the wrong
thing to send.

\textit{Why a soliton.---}The two effects that balance in a real soliton both have
counterparts here.
The derivative term $|\dot a|^2$ in Eq.~(\ref{eq:energy}) charges the drive for fast
features exactly as dispersion does, while the objective $|a|^{2n}$ rewards
concentration exactly as self-focusing does. The decay rate $\kappa$ sets the exchange
rate, and the optimum is the profile where the two balance.

The balance is quantitative.  {The fixed quantity in
Eq.~(\ref{eq:energy})} splits in a
ratio set by the response order alone---the derivative term takes the fraction
$(n-1)/2n$ and the storage term $(n+1)/2n$~\cite{SM}---independent of $\kappa$,
$\kappa_e$, $E$, and every material constant. As $n\to\infty$ the split approaches
equipartition and the stored field approaches the two-sided exponential $e^{-\kappa|t|}$:
the ringdown joined to its own time reverse at the peak [Fig.~\ref{fig:exact}(b)]. The
stored intensity narrows with order but saturates at the full width $\ln2/\kappa$ set
by the resonance itself; what keeps sharpening is the nonlinear source $|a|^{2n}$,
whose width falls off as $1/(n\kappa)$~\cite{SM}.

The input that sustains this field is asymmetric [Fig.~\ref{fig:exact}(c)]. It rises as
$e^{\kappa t}$ and falls as $e^{-(2n-1)\kappa t}$, remaining on and in phase after the
stored peak.  {In the language of complex-frequency excitation, where a
drive $\propto e^{\gamma t}$ on carrier $\omega_0$ addresses the resonator at the
single point $\omega_0+i\gamma$ of the complex-frequency
plane~\cite{Baranov2017,Kim2025Review}, the optimal input occupies no such point: its
local rate $d\ln|s_{\rm opt}|/dt=-\kappa\left[(n-1)+n\tanh x\right]$ sweeps
continuously from $+\kappa$ on the leading tail to $-(2n-1)\kappa$ on the trailing
tail~\cite{SM}. The optimum is a trajectory across the complex-frequency plane, of the
kind chirped pulses are built to follow~\cite{KrasnokSeletskiy2026}, and any
single-rate exponential---the time-reversed drive included---samples one point of it.
The suboptimality of the rising exponential, quantified next, is the cost of that
single point.} Its
root-mean-square bandwidth grows only as $\kappa\sqrt{2n/3}$~\cite{SM}, far slower than
the trailing rate, because little energy sits in the fast trailing edge.

\textit{Practical drives.---} {The equality waveform is smooth and
two-sided, and no laboratory source emits it directly. What matters for an experiment
is how much of the bound survives with pulses that can actually be synthesized.} For
the one-sided rising
exponential $s\propto e^{\gamma t}$ switched off at $t_0$, the yield is
 {available in closed form}~\cite{SM} and is maximized at exactly
$\gamma^*=(n-1)\kappa/n$: always below
$\kappa$, half of it at $n=2$. The time-reversal rate $\gamma=\kappa$ maximizes the
stored energy at the switching instant, which answers the linear question, while
conversion integrates intensity over all time and prefers a slower rise. Even at its
best rate the exponential retains a fraction of the bound that falls from $79.0\%$ at
$n=2$ to $2/e\simeq73.6\%$ as $n\to\infty$ [Fig.~\ref{fig:practical}(a)]. The ideal
switch-off is also unphysical: the step gives the pulse unbounded root-mean-square
bandwidth.

 {A second, finite fall rate removes both problems.} The continuous
two-rate pulse
$s\propto e^{\gamma(t-t_0)}$ for $t<t_0$ and $e^{-\beta(t-t_0)}$ after echoes the two
tails of the exact solution, and finite switching raises the yield rather than costing
it. With the rise reoptimized at each $\beta$, the retained fraction at $n=2$ climbs
from $79.0\%$ in the step limit to $97.7\%$ at the joint optimum
$(\gamma,\beta)=(0.756,1.349)\kappa$, and across integer $2\le n\le20$ the optimized
two-rate pulse holds $97.45$--$97.68\%$ of the bound
[Fig.~\ref{fig:practical}(a)]---numerical evidence over that interval rather than a
proved constant. The cost is bandwidth: the optimal switching rate grows as
$\beta\simeq n\kappa$ [Fig.~\ref{fig:practical}(b)], so the modulator must pass roughly
$n$ linewidths, and a pulse switched at only $\beta=\kappa$ keeps a mere $19\%$ at
$n=10$~\cite{SM}.

Symmetric transform-limited pulses fare worse the higher the order: against
width-optimized Gaussian and hyperbolic-secant pulses of the same energy, the exact
waveform yields over a third more at $n=5$ and about three times more at $n=10$
[Fig.~\ref{fig:practical}(a)]; Fig.~S1(a) of the Supplemental Material~\cite{SM} compares their $n=5$ envelopes.

\textit{Coupling and the generated channel.---} {The bound has so far
been read at fixed device parameters, with the waveform as the only free choice. Two of
those parameters are in fact design variables, and the bound tells us where to set
them: the strength of the input coupler, and the linewidth of the channel that collects
the generated light.} Equation~(\ref{eq:bound}) depends on the
coupler only through $\rho^n/\kappa=\kappa_e^n/(\kappa_e+\kappa_i)^{n+1}$, so when the
intrinsic rate is fixed and the input coupling is adjustable---an evanescent gap, an
antenna---the bound itself can be maximized:
\begin{equation}
 \boxed{\ \kappa_e^{\rm opt}=n\,\kappa_i\ }
 \label{eq:coupling}
\end{equation}
Critical coupling is the $n=1$ case. An $n$th-order device should be overcoupled
$n$-fold, that is, loaded $n$ times harder than the reflectionless condition, because
the yield rewards stored intensity more steeply than it penalizes the shortened
lifetime. The waveform must be reoptimized along a coupling scan, since $\kappa$ grows
with $\kappa_e$. The gain over critical coupling is modest at low order ($32/27$ at
$n=2$), so a tunable coupler tests this better than a set of fixed gaps
(Fig.~S3~\cite{SM}).

 {Everything so far rests on one assumption still to be tested: that the
generated channel is spectrally flat, so that the emitted energy follows $U_n$. We now
lift that idealization, both to say how broad the channel must actually be and to cover
devices in which it is narrow.} Let the generated field occupy its own
resonant mode $b$ with loaded rate $\kappa_b$, output coupling $\kappa_{b,e}$, and
detuning $\Delta_b$ from the emission line, driven by the nonlinear source as
$\dot b=-(\kappa_b+i\Delta_b)b-iga^n$ with $g$ the nonlinear coupling constant.
 {Solving for $b$ and integrating the outflow,} the
emitted energy is a filtered version of the source spectrum,
\begin{equation}
 U_{\rm out}=\frac{2\kappa_{b,e}|g|^2}{2\pi}\int
 \frac{|\widetilde{a^n}(\Omega)|^2}{\kappa_b^2+(\Delta_b-\Omega)^2}\,d\Omega ,
 \label{eq:filtered}
\end{equation}
where $\widetilde{a^n}$ is the Fourier transform of $a^n$ and $\Omega$ the offset from
the generated resonance. When $\kappa_b$ far exceeds the source bandwidth
$\sigma_{\Omega,a^n}=n\kappa\sqrt{(n-1)/(3n-1)}$, Eq.~(\ref{eq:filtered}) reduces to
$U_n$ and everything above applies; measured in units of $\sigma_{\Omega,a^n}$ the
retention is weakly order dependent, and $\kappa_b\gtrsim5\sigma_{\Omega,a^n}$ keeps
the correction below about $4\%$ (Fig.~S4(a)~\cite{SM}). In the opposite limit---a
generated resonance much narrower than the source and tuned to it---only the
zero-frequency component of $a^n$ survives, $U_{\rm out}\propto|\int a^ndt|^2$, and the
optimum changes character: for $n>2$ the best stored field is the much broader soliton
$\mathrm{sech}^{2/(n-2)}[\tfrac{n-2}{2}\kappa(t-t_0)]$, and the coupling law softens to
$\kappa_e^{\rm opt}=(n/2)\kappa_i$ (Fig.~S3(b,c)~\cite{SM}). The output channel decides
which law a device obeys.

 {These two limits are the extremes of a continuum, and a real device
sits between them, with a measured output response that is neither flat nor a bare
Lorentzian. To reach such a device we drop every closed-form assumption and ask only
what any optimal input must satisfy.} With $a=\mathcal Ls$ the linear pump
response and $\mathcal K$ the full generated-side transfer---cavity response, phase
matching, propagation, collection---every input maximizing $\|\mathcal K(a^n)\|_2^2$ at
fixed energy satisfies the stationarity condition
\begin{equation}
 \mathcal L^\dagger\!\left[(a^*)^{n-1}\mathcal K^\dagger\mathcal K(a^n)\right]
 =\lambda s ,
 \label{eq:adjoint}
\end{equation}
with $\lambda$ a Lagrange multiplier and $\dagger$ the adjoint. This is a stationarity
condition rather than a bound; it supports a fixed-point iteration for realistic
devices, including measured modulator responses~\cite{Baxter2023}, and it reduces to
the closed forms above for a one-pole pump with a flat or narrow output. Multiwave
mixing with independently shaped pumps collapses onto the same solution: when the pump
resonances share a loaded linewidth, H\"older's inequality forces every field into the
profile of Eq.~(\ref{eq:aopt}) with $n$ the total number of participating pump
photons~\cite{SM}.

\textit{Self-action.---} {Both the bound and the stationarity condition
treat the resonator as a fixed linear object. That assumption is the one most at risk
from the result itself, because the waveform the bound selects concentrates the stored
intensity.} At the energies the bound rewards, the resonator's own
parameters move: in a dielectric cavity the stored intensity shifts the refractive
index, which drags the resonance frequency. The bound is exactly indifferent to the
reactive part of this response. Let the resonance shift linearly
with stored energy,
$\dot a=-[\kappa+i(\Delta_0+K|a|^2)]a+\sqrt{2\kappa_e}\,s$, where $K$ is the Kerr shift
per unit stored energy and $\Delta_0$ the static detuning. Substituting $a=qe^{-i\Phi}$
with $\dot\Phi=\Delta_0+K|q|^2$ removes the shift entirely:
\begin{equation}
 \dot q=-\kappa q+\sqrt{2\kappa_e}\,e^{i\Phi}s ,
 \label{eq:gauge}
\end{equation}
with $|a|=|q|$ and the incident energy untouched. The optimal envelope is therefore the
same soliton at any Kerr strength for which the fixed-mode, memoryless single-pole
model and the required phase bandwidth hold, and the optimal drive differs only by a
closed-form phase: its instantaneous frequency tracks the power-shifted resonance. For
$n=2$ the chirp integrates to
$\Phi_{\rm Kerr}=(KA_2^2/\kappa)\tanh[\kappa(t-t_0)]$. Self-phase
modulation, the leading way strong pumping alters a dielectric resonator, therefore
costs no yield within the single-mode model; it re-chirps the drive. A narrow generated
channel does see the induced phase, through $a^n=q^ne^{-in\Phi}$, and its full filtered
response must be evaluated~\cite{SM}.

Phase tracking cannot remove nonlinear dissipation. Two-photon and free-carrier
absorption enter as an intensity-dependent $\kappa$ and remain constrained
perturbatively, with the first-order yield correction evaluated on the unperturbed
optimum and the requirement that nonlinear loss stay small against $\kappa$ kept
 {as a validity condition}~\cite{SM}. Slow thermal drift is quasi-static
and disappears into
calibration, with $\omega_0$ and $\kappa$ measured at the operating power; carrier
memory requires an augmented nonlinear model beyond Eq.~(\ref{eq:adjoint}).
Stationarity makes the yield quadratically insensitive to small feasible waveform
errors, but not to arbitrary parameter errors: energy mismatch, nonlinear loss, and
changes of the device response all perturb it at first order~\cite{SM}.

\textit{Photon-pair sources.---} {The analysis so far has been classical
and stimulated. It carries over to spontaneous emission of photon pairs, where the
quantity being maximized is a probability and the bound becomes a ceiling on source
brightness.} At low gain the same functional governs spontaneous
processes,  {with one change in what the exponent counts}. If
 {a single conversion event} annihilates $m$ pump
photons, first-order perturbation theory with a broadband generated channel gives a
generation probability proportional to $\int|a|^{2m}dt$: everything above applies with
$n=m$, and the exponent counts pump photons destroyed rather than photons created.
Degenerate spontaneous four-wave mixing consumes two pump photons per pair, so the
$n=2$ soliton maximizes broadband pair-generation probability at fixed pulse energy and
the pump coupler belongs at $\kappa_e=2\kappa_i$---consistent with the finding that
optimal microring pair sources sit above critical coupling~\cite{Lukens2026}.
Spontaneous parametric down-conversion consumes one pump photon and so has no
finite-duration optimum, however many photons it creates~\cite{Corona2011}.  {Brightness
does not imply state purity:} heralded purity lives in the joint spectral amplitude,
the two-photon spectral wave function whose factorability sets the state quality, and
the brightness-optimal asymmetric pump does not in general factorize
it~\cite{Grice2001,Christensen2016,Rodda2024}. Mapping that trade-off is a separate,
constrained problem.

\textit{Outlook.---} {The predicted differences are large---the best
rising exponential loses $21\%$ of the bound at $n=2$, and a width-optimized Gaussian
loses a factor of three at $n=10$---so the waveform comparison survives percent-level
systematic error. The most direct test uses a
pump-resonant device with a broad generated channel, such as} a periodically poled
lithium niobate racetrack with a strongly overcoupled
second-harmonic port. At $1550$~nm and loaded $Q=10^6$ the linewidth is
$\kappa/2\pi=96.7$~MHz; the optimized $n=2$ two-rate pulse needs rates
$(\gamma,\beta)/2\pi=(73,130)$~MHz and, at $E=10$~pJ, a peak power of
$5.9$~mW---nanosecond waveforms synthesized by an electro-optic in-phase/quadrature
modulator and an arbitrary-waveform generator, with no femtosecond apparatus. The
nonlinear coefficient never enters the comparison: measure $\kappa_e$, $\kappa_i$, and
the detuning by linear spectroscopy, equalize pulse energies shot by shot, interleave
candidate waveforms against thermal drift, and compare yields~\cite{SM}.
Superconducting circuits invert the difficulty: the same input--output equation governs
circuit quantum electrodynamics~\cite{Gardiner1985,Blais2021}, multiphoton and
Josephson processes supply the exponent, and narrow linewidths paired with electronics
of usable bandwidth $B$ leave a large switching margin $B/[n\kappa/(2\pi)]$, making
microwave resonators an accessible place to test Eq.~(\ref{eq:bound}).

What remains open is where the closed form stops. Equation~(\ref{eq:adjoint}) turns
real devices---finite output linewidths, phase-matching windows, measured
modulators---into a computable problem whose global optimality is unproven. The
brightness--purity front of pulsed pair sources is a constrained version of the same
question. And once conversion becomes efficient, the pump equation itself turns
nonlinear through depletion, critical-power and nonlinear impedance-matching conditions
take over~\cite{Rodriguez2007}, and the fate of the soliton optimum is unknown.

\textit{Conclusion.---}At fixed pulse energy the optimal drive for $n$th-order resonant
conversion is now known in closed form: the stored field is the soliton
$\mathrm{sech}^{1/(n-1)}[(n-1)\kappa t]$, the input that builds it rises as
$e^{\kappa t}$ and falls as $e^{-(2n-1)\kappa t}$, the yield cannot exceed
$\mathcal C_n\rho^nE^n/\kappa$, and the coupler belongs at $\kappa_e=n\kappa_i$. The
response order and the loaded linewidth fix all of it, so linear spectroscopy suffices
to score a real device against the bound, and two exponentials joined at a peak already
reach $97\%$ of it. Because the optimum follows from an optimization rather than from
material dynamics, it carries over unchanged from a microring to a superconducting
circuit.

\FloatBarrier
\textit{Data and software availability.}---The code and numerical data supporting the
reported results are available from the author upon request.

\begin{acknowledgments}
The author acknowledges financial support from the U.S. Department of Energy through
Brookhaven Science Associates, LLC, under Subcontract No.~463609 and Prime Contract
No.~DE-SC0012704, and from the U.S. Air Force Office of Scientific Research (AFOSR)
under Grant No.~FA9550-26-1-B070.
\end{acknowledgments}

\nocite{Weiner2011,Corkum1993,Lewenstein1994,Lalanne2018,Chong2010}
\bibliographystyle{apsrev4-2}
\bibliography{refs}

\end{document}